\documentclass[prl,aps,
superscriptaddress,tightenlines,twocolumn,preprintnumbers]{revtex4}

\usepackage{graphicx}
\usepackage{amssymb}

\newcommand{\beq}{\begin{equation}}
\newcommand{\eeq}{\end{equation}}
\newcommand{\bqa}{\begin{eqnarray}}
\newcommand{\eqa}{\end{eqnarray}}
\newcommand{\be}{\begin{equation}}
\newcommand{\ee}{\end{equation}}
\newcommand{\bea}{\begin{eqnarray}}
\newcommand{\eea}{\end{eqnarray}}
\newcommand{\nn}{\nonumber}
\newcommand{\0}{\over }

\newcommand{\6}{\partial }

\begin{document}

\title{Plasma Instabilities in an Anisotropically Expanding Geometry}

\preprint{BI-TP 2006/11}
\preprint{TUW-06-04}

\author{Paul Romatschke}
\affiliation{Fakult\"at f\"ur Physik, Universit\"at Bielefeld,
D-33501 Bielefeld, Germany}
\author{Anton Rebhan}
\affiliation{Institut f\"ur Theoretische Physik, Technische Universit\"at Wien,
        Wiedner Hauptstrasse 8-10, A-1040 Vienna, Austria}


\begin{abstract}
We study (3+1)D kinetic (Boltzmann-Vlasov)
equations for relativistic 
plasma particles in a one-dimensionally
expanding geometry (a special Kasner-type universe)
motivated by ultrarelativistic heavy-ion
collisions.
We set up local equations in terms of Yang-Mills
potentials and auxiliary fields 
that allow simulations of hard-(expanding)-loop (HEL) dynamics on a lattice.
We determine numerically the evolution of plasma instabilities in the linear
(Abelian) regime and also derive their late-time behavior analytically,
which is consistent with recent numerical results on
the evolution of the so-called melting color-glass condensate.
We also find a significant delay in the onset of growth of
plasma instabilities which are triggered by small rapidity fluctuations,
even when the initial state is highly anisotropic.
\end{abstract}

\maketitle

Plasma instabilities have recently been suggested to play a major role
in the equilibration of matter created by an ultrarelativistic
heavy-ion collision, e.g. at the Relativistic Heavy Ion Collider
(RHIC) or the Large Hadron Collider (LHC)
\cite{Randrup:2003cw,Romatschke:2003ms,Arnold:2003rq,Rebhan:2004ur,Dumitru:2005gp,Arnold:2005vb}.
Shortly after such a collision, saturation scenarios \cite{McLerran:1993ni,Iancu:2003xm} indicate that
the typical momentum of a particle in the local plasma rest frame
is much larger transverse to the collision axis than  
parallel to it. 
This momentum
anisotropy inevitably leads to 
a so-called Weibel \cite{Weibel:1959} (or
filamentation) plasma instability that manifests itself by rapidly
growing transverse magnetic fields. If large enough, a transverse
magnetic field bends the trajectories of particles out of 
the transverse plane, thus making the system more isotropic.
Plasma instabilities
are therefore a prime candidate for causing rapid
isotropization of a quark-gluon
plasma, especially since they act on a time-scale that is
parametrically shorter than that of scatterings by at least one
power of the strong coupling constant $g$.

There are, however, some caveats: for instance, 
numerical simulations \cite{Rebhan:2004ur,Arnold:2005vb}
of non-Abelian Yang-Mills equations in the 
(stationary) hard-loop approximation
\cite{Mrowczynski:2004kv}
have shown that the initially exponential
growth of gauge fields slows down to a weak 
linear growth when self-interactions of the unstable
modes become important and energy in unstable modes cascades
to stable modes of higher momentum \cite{Arnold:2005ef}.
Interestingly, this may lead to 
the generation of anomalously low viscosity \cite{Asakawa:2006tc}.

Maybe more importantly, the matter created in a heavy-ion collision is
believed to escape relatively unimpeded in the longitudinal direction 
(the direction of the collision axis). This effectively 
one-dimensional expansion decreases the density of
hard particles \cite{Mueller:1999pi,Baier:2002bt} and thus attenuates the growth rate of
plasma instabilities.
On the other hand, the expansion increases the
degree of anisotropy, thus making more and more higher-momentum modes unstable.
While numerical simulations of classical Yang-Mills
dynamics have provided some qualitative understanding of the
counterplay of plasma instabilities and expansion
\cite{Romatschke:2005pm}, 
an analysis based
on the hard-loop approximation is desirable,
in particular to address systematically the fate of
non-Abelian plasma instabilities and of the associated energy cascade.
The purpose of this Letter
is to develop the basis for such a treatment.
We generalize the anisotropic hard-loop effective theory of Refs.~\cite{Rebhan:2004ur,Arnold:2005vb}
to the case of a dynamical, boost-invariantly expanding background,
thus preparing the ground for corresponding lattice simulations.
We work out explicitly the already rather nontrivial dynamics in the
linear (Abelian) regime (which thus are in principle of interest
also to ultrarelativistic conventional plasma physics)
and discuss possible implications for ultrarelativistic heavy-ion collisions.

Ignoring the effects of collisions, the dynamics of collective
modes in a non-Abelian plasma
is determined by the gauge covariant Boltzmann-Vlasov equations
\be\label{vDf}
p\cdot D\, \delta f_a(\1p,\1x,t)=g p_\mu F^{\mu\nu}_a \6^{(p)}_\nu f_0(\1p,\1x,t),
\ee
coupled to the Yang-Mills equations
\be\label{DFj}
D_\mu F^{\mu\nu}_a=j^\nu_a=
g \int{d^3p\0(2\pi)^3} \frac{p^\mu}{2p^0} \delta f_a(\1p,\1x,t).
\ee
Here $F^{\mu\nu}_a=\partial^\mu A^\nu_a-\partial^\nu A^\mu_a+g
f_{abc} A^\mu_b A^\nu_c$ is the field strength tensor in the adjoint
representation, $D^\mu_{ac}=\delta_{ac} \partial^\mu+ g f_{abc}
A^\mu_b$ is the gauge-covariant derivative,
$A^\mu$ is the
gauge-potential. $f_0(\1p,\1x,t)$ 
is the (suitably normalized)
color-neutral background distribution of hard particles
and
$\delta f_a(\1p,\1x,t)$ its colored fluctuations.
Eq.(\ref{vDf}) requires that the background $f_0$
satisfies $p\cdot \6\, f_0(\1p,\1x,t)=0$ (which is trivially fulfilled
when $f_0$ is space-time independent). Assuming that the matter
created after a heavy-ion collision
expands longitudinally in a boost-invariant way 
and is sufficiently large in the transverse direction such that
transverse gradients are small, we take
\beq
f_0(\mathbf p,x)=f_0(p_\perp,p^z,z,t)=f_0(p_\perp,t p^z-z p^0)
\eeq
which also satisfies $p\cdot \6\, f_0(\1p,\1x,t)=0$.

To describe the dynamics of fluctuations
around a boost-invariant background, convenient 
coordinates are proper time $\tau=\sqrt{t^2-z^2}$ and space-time
rapidity $\eta={\rm arctanh}\frac{z}{t}$. The metric in these
coordinates is $ g_{\tau \tau}{\rm =1}$, $g_{ij}{\rm =-}\delta_{ij}{\rm
}$, $g_{\eta\eta}{\rm =-}\tau^2$,
$i{\rm =}1,2$, 
thus corresponding to a one-dimensionally expanding space-time geometry.

Transforming the set of equations (\ref{vDf},\ref{DFj}) to the
coordinates $x^\alpha=(\tau,x^i,\eta)$ (denoted collectively by Greek
letters from the beginning of the alphabet) we find
\bqa
p \cdot D \,\delta f^a|_{p^\mu}
&=&g p^\alpha F_{\alpha\beta}^a  \6_{(p)}^\beta
f_0(\1p_\perp,p_\eta), \label{transeq}\\
{1\0\tau}
D_\alpha \left(\tau 
F^{\alpha\beta}
\right)&=& j^{\beta}=
\frac{g}{2} \int{d^2p_\perp dy\0(2\pi)^3} p^\beta
\delta f\, ,
\label{maxweq}
\eqa
where the derivative of $\delta f^a$ has to be taken at fixed $p^\mu$
as opposed to fixed $p^\alpha$.
Here we have introduced
also momentum rapidity $y={\rm atanh}(p^0/p^z)$, 
such that 
$p^\tau=|{\bf p^\perp}| \cosh{(y-\eta)}$,
$p^\eta=|{\bf p^\perp}| \tau^{-1} \sinh{(y-\eta)}$.

We assume that at the hypersurface $\tau=\tau_{\rm iso}$ the
background $f_0$ is isotropic and choose the specific model
\be\label{faniso}
f_0(\mathbf p,x)
=f_{\rm iso}\left(\sqrt{p_\perp^2+p_\eta^2/\tau_{\rm iso}^2}\right),
\ee
which corresponds to increasingly oblate momentum space anisotropy 
at $\tau>\tau_{\rm iso}$
(but prolate anisotropy for $\tau<\tau_{\rm iso}$). 
However, in what follows we shall start the time evolution at
nonzero proper time $\tau_0$, allowing for the fact that a plasma
description will not make sense at arbitrarily small times,
and we shall mostly consider the situation
that the initial momentum distribution is highly oblate
at $\tau_0$, i.e. $\tau_0
\gg \tau_{\rm iso}$.
The distribution function $f_0$ has the same form as the one used in 
Refs.~\cite{Romatschke:2003ms,Rebhan:2004ur},
but the anisotropy parameter $\xi$ therein is now space-time dependent
according to $\xi(\tau)=(\tau/\tau_{\rm iso})^2-1.$ 

Since $p\cdot\6 \,[\6_{(p)}^\alpha f_0(\mathbf p_\perp,p_\eta)]=0$, we 
can solve Eq.~(\ref{transeq}) by
introducing
auxiliary fields $W^a_{\alpha}(\tau,x^{i},\eta;v^i,y)$ in
\be
\delta f^a(x;p)=-g W^a_\alpha(\tau,x^{i},\eta;v^i,y)
\6_{(p)}^\alpha f_0(p_\perp,p_\eta)
\ee
that obey
\be\label{VDW}
v\cdot D\, W_\alpha(\tau,x^{i},\eta;v^i,y)=v^\beta F_{\alpha\beta},
\ee
where $v^\alpha\equiv {p^\alpha\0 |{\mathbf p_\perp}|}
=(\cosh(y-\eta),\cos\phi, \sin\phi,\frac{\sinh(y-\eta)}{\tau}).$
At any given space-time point the fields $W_\alpha$ depend only on the
velocity of the hard particles and not on their momentum scale, 
and thus directly
generalize the auxiliary fields $W_\mu(x;\mathbf v)$ of
the hard-loop formalism in a static background distribution 
$f_0$ \cite{Blaizot:2001nr,Mrowczynski:2004kv}.

For $f_0$ of Eq.(\ref{faniso}), the induced
current $j^\beta$ takes the form
\be
j^\alpha\!=\!-{m_D^2\02} \int_0^{2\pi} {d\phi \0 2\pi}
\int
\frac{dy\ v^\alpha}
{\left(1+{v_\eta^2\0\tau_{\rm iso}^2}\right)^{2}} 
\left\{v^j W_j
-{v_\eta\0\tau_{\rm iso}^2} W_\eta\right\}\, ,
\label{current}
\ee
where $m^2_D=- g^2 \int_0^\infty {dp\,p^2\0(2\pi)^2} f'_{\rm iso}(p)$
is the Debye mass of the isotropic case.
A short calculation confirms that the current is covariantly conserved,
$D_\alpha \tau j^\alpha=0$.

The Eqs. (\ref{VDW}), (\ref{current}) together with the Yang-Mills
equations (\ref{maxweq}) can be simulated on a 3-dimensional
lattice by discretizing space-time and introducing lattice links in a
standard way \cite{Krasnitz:1998ns}
and by also discretizing the residual momentum variables $\phi$ and $y$
in order to have a finite number of $W$ fields.
This directly generalizes the discretized
hard-loop effective equations of motions of Ref.\cite{Rebhan:2004ur} to
what may be called the hard-expanding-loop (HEL) case.
As in the stationary case, 
momentum discretizations that respect reflection invariance 
(now with respect to $\phi$ and $y-\eta$)
automatically ensure 
covariant current conservation, $D_\alpha \tau j^\alpha=0$.

In this Letter, we initiate this program by studying
the onset of plasma instabilities
and their evolution in the linear regime, where
their non-Abelian self-interactions are still
negligible and where we can avoid a discretization of $\phi$ and $y$
by solving the equations of motions of the auxiliary fields
$W_\alpha(\tau,x^{i},\eta;\phi,y)$.
In a stationary plasma with oblate momentum-space anisotropy, the
most unstable modes have wave vectors along the longitudinal direction.
A particularly interesting case can thus be studied
by neglecting the transverse dynamics ($\partial_i A^\alpha=0$).
Linearizing in the gauge potentials we have
\beq
\left[
\tau^{-1}\partial_\tau \tau \partial_\tau-\tau^{-2} \partial_\eta^2
\right] A^{i}(\tau,\eta)=j^{i},\quad \tau\partial_\tau \tau^{-1}
\partial_\tau A_\eta = j_\eta,
\label{Meq}
\eeq
in the gauge
$A^\tau=0$. Eq.~(\ref{VDW}) can be solved by the method of
characteristics which gives
\bea
W_{\alpha}(\tau,\eta;\phi,y)&=&\int_{\tau_{0}}^{\tau} d\tau^\prime 
\frac{
\left. v^\beta F_{\alpha\beta} \right|_{\tau^\prime,\eta(\tau^\prime)}
}{\cosh(y-\eta(\tau^\prime))}\, ,\nn\\
y-\eta(\tau^\prime)&=&{\rm asinh}{\left(\frac{\tau}{\tau^\prime} 
\sinh{(y-\eta)}\right)},
\eea
where $\eta{(\tau^\prime)}$ is the solution along the characteristic.
Within our approximation we can thus proceed to evaluate the $d\phi$ integral
in Eq.(\ref{current}), finding
\bea
j^{i}\!&=&\!-\frac{m_D^2}{4}\,
\int_{-\infty}^\infty dy \,
\left(1+{v_\eta^2\0\tau_{\rm iso}^2} \right)^{-2} 
\int^\tau_{\tau_{0}} d\tau^\prime 
 \nn\\
&&
\!\!\!\!\!\!\!\!\!\!\!\!
\times\left[\left(\partial_\tau^\prime-\frac{
\tanh{\bar{\eta}^\prime}
}{\tau^\prime}\partial_{\eta^\prime}
\right) A^{i}(\tau^\prime,\eta^\prime)
+{v_\eta\0\tau_{\rm iso}^2}\,
\frac{\partial_{\eta^\prime}
A^{i}(\tau^\prime,\eta^\prime)}{\cosh\bar{\eta}^\prime}\right],\nn\\
j^\eta\!\!&=&\!\!-{m_D^2 \0 2\tau_{\rm iso}^2}
\int
\frac{dy \,v^\eta v_\eta}
{\left(1+{v_\eta^2\0\tau_{\rm iso}^2} \right)^{2}} 
\int^\tau_{\tau_{0}} d\tau^\prime 
\partial_{\tau^\prime}
A_{\eta}(\tau^\prime,\eta^\prime),
\label{jicur}
\eea
where $\eta^\prime=\eta(\tau^\prime)$ and 
$\bar{\eta}^\prime=\eta(\tau^\prime)-y$. 

Introducing a Fourier transform in space-time rapidity,
\be
A^{i}(\tau,\eta)=\int \frac{d \nu}{2 \pi} \exp(i \nu \eta)
\widetilde{A}^{i}(\tau,\nu),
\ee
and choosing $A^\mu(\tau_0)=0$ for simplicity,
we find
\bea
&&\widetilde{j}^{i}(\tau,\nu)=-\frac{m_D^2}{4}\int \frac{dy}
{\left(1+{\tau^2\sinh^2{y}\0\tau_{\rm iso}^2} \right)^{2}} 
\biggl\{\widetilde{A}^{i}(\tau,\nu)
\nn\\
&&\quad-\int_{\tau_0}^\tau d\tau^\prime 
\frac{\widetilde{A}^{i}(\tau^\prime,\nu) \tau^{\prime 2}}{\tau_{\rm
iso}^2}
\partial_{\tau^\prime} e^{i\nu\left[y-{\rm
asinh}{\left(\frac{\tau}{\tau^\prime} \sinh{y}\right)}\right]}
\biggr\},\nn\eea
\bea
&&\widetilde{j}^{\eta}(\tau,\nu)={m_D^2 \0 2\tau_{\rm iso}^2}
\int \frac{dy \,\sinh^2{y}}
{\left(1+{\tau^2\sinh^2{y}\0\tau_{\rm iso}^2} \right)^{2}} 
\biggl\{\widetilde{A}_{\eta}(\tau,\nu)
\nn\\
&&\quad-\int_{\tau_0}^\tau d\tau^\prime 
\widetilde{A}_\eta(\tau^\prime,\nu)
\partial_{\tau^\prime}
e^{i\nu\left[y-{\rm
asinh}{\left(\frac{\tau}{\tau^\prime} \sinh{y}\right)}\right]}
\biggr\}.\qquad
\label{ecurr}
\eea

Below we solve
the integro-differential Eqs.\ (\ref{Meq}), (\ref{ecurr}) 
numerically. 
The late time behavior $\tau\gg \tau_0$, however,
may be studied analytically by expanding $\exp{\left[i\nu\left(y-{\rm
asinh}{\left(\frac{\tau}{\tau^\prime} \sinh{y}\right)}\right)\right]}$
in Eq.(\ref{ecurr}) around $\tau^\prime=\tau$ and subsequently acting
with $\partial_{\tau}^2 \tau^2$ on Eqs.(\ref{Meq}).
In this limit, the integro-differential equations turn into ordinary
differential equations for each mode $\nu$,
\bea
&\left[
\partial_\tau^2 \tau \partial_\tau \tau \partial_\tau+
\nu^2\partial_\tau^2
+\mu \partial_\tau^2 \tau
-\mu \nu^2{\tau^{-1}}
\right] \widetilde{A}^{i}(\tau,\nu) \simeq 0,&\label{apEQ}\qquad\\
&\left[
\partial_\tau {\tau^{-1}} \partial_\tau
+2\mu{\tau^{-2}}
\right] \widetilde{A}_{\eta}(\tau,\nu) \simeq 0,&
\label{AetaEQ}
\eea
where $\mu={1\08}{m_D^2 \pi \tau_{\rm iso}}$. From Eq.~(\ref{AetaEQ}) we
find for the late-time behavior of 
longitudinal fields
$A_z=\tau^{-1} A_\eta\cosh\eta$ 
\beq
\tau^{-1}\widetilde{A}_\eta(\tau,\nu) \simeq c_1 J_2\left(
2 \sqrt{2 \mu \tau}\right)+
c_2 Y_2\left(2 
\sqrt{2\mu\tau}\right),
\eeq
where $J_n(x)$ and $Y_n(x)$ are Bessel functions of the first and second kind,
respectively, and $c_{1,2}$ are
constants. The asymptotic behavior of these functions is 
oscillatory with amplitude $\sim x^{-1/2}$, so $\tau^{-1}A_\eta$ 
only has stable modes.

For very infrared modes $\nu\ll 1$, the terms proportional to $\nu$
in Eq.(\ref{apEQ}) can be neglected and we find for $A^i$ 
\be
\widetilde{A}^{i}(\tau,\nu\!\ll\!1) \simeq c_1 J_0 \left(2
\sqrt{\mu\tau}\right)+ c_2 
Y_0 \left(2
\sqrt{
\mu\tau
}\right),
\label{osc}
\ee
which again correspond to stable oscillatory solutions, but whose
frequency is a factor $\sqrt2$ smaller than that of $\widetilde{A}_\eta$.
This is consistent with the analytic results
of Refs.~\cite{Romatschke:2003ms,Rebhan:2004ur}
on nonexpanding anisotropic plasmas, 
where the ratio of the longitudinal
and the transverse plasma frequency was also found to approach $\sqrt2$ in
the limit of infinite anisotropy parameter $\xi$.

For high-momentum modes
$\nu\gg 1$, however, only the terms proportional to $\nu$ in
Eq.(\ref{apEQ}) matter, and we find 
\be
\widetilde{A}^{i}(\tau,\nu\!\gg\!1) \simeq
c_1 \sqrt{\tau}
I_1 \left(2
\sqrt{
\mu\tau
}
\right)+ c_2 
\sqrt{\tau}
K_1\left(2
\sqrt{
\mu\tau
}
\right),
\label{bound}
\ee
where $I,K$ are modified Bessel functions with asymptotic behavior 
$I_n(x)\simeq\frac{\exp(x)}{\sqrt{2 \pi x}}$, 
$K_n(x)\simeq\frac{\exp(-x)}{\sqrt{2 x/\pi}}$. 
Clearly, the first of these solutions corresponds to a rapidly growing
mode which leads us to expect that the large $\nu$ modes of
$A^i$ will be the dominant modes at sufficiently late times with
a behavior of
\be
\widetilde{A}^{i}(\tau) \sim \tau^{1/4} \exp\left(2
\sqrt{
\mu\tau
}
\right).
\label{asymbeh}
\ee
This behavior has qualitatively been found by numerical simulations of
the melting color-glass condensate \cite{Romatschke:2005pm}.
For moderate momenta $\nu\sim 1$ the solutions to Eq.(\ref{apEQ}) are more
complicated and can be given in terms of 
generalized hypergeometric functions $_2F_3$ and a Meijer G-function. 
The dominant contribution turns out to be
\be
\widetilde{A}^{i}(\tau,\nu)\sim \tau\, \,_2F_3
\left(\textstyle{\frac{3-s}{2}},\textstyle{\frac{3+s}{2}};2,2-i
\nu,2+i \nu;-\mu\tau
\right)
\label{intermediate}
\ee
with $s=\sqrt{1+4 \nu^2}$,
which interpolates between the simple cases $\nu\ll1$ (Eq.~(\ref{osc})) and
$\nu\gg 1$, Eq.~(\ref{asymbeh}).

Information on the behavior of $A^i$ at very early times can be gained by
studying Eq.(\ref{ecurr}) for $\nu \gg \tau/\tau_{\rm iso}$. Expanding around
the stationary point $y=0$, 
we find the oscillatory behavior $\widetilde{A}^{i}(\tau,\nu) \simeq c_1
\tau^{i \nu}$.

In order to investigate the onset of plasma instabilities
we have solved
the non-approximated integro-differential Eqs.\
(\ref{Meq}), (\ref{ecurr}) numerically. 
To this end 
we introduce $\widetilde{\Pi}^{i}(\tau,\nu)=\tau \partial_\tau
\widetilde{A}^{i}(\tau,\nu)$ which obeys
$
\partial_\tau \widetilde{\Pi}^{i}(\tau,\nu)=-{\nu^2}{\tau^{-1}} 
\widetilde{A}^{i}(\tau,\nu)+\tau\widetilde{j}^{i},
$ 
and apply a leap-frog algorithm to solve the coupled equations after
discretizing both the variable $\tau$ and $\tau'$
of the memory integral in
$\widetilde{j}^{i}$,
with initial conditions $\widetilde{A}^{i}(\tau_0,\nu)=0$,
$\widetilde{\Pi}^{i}(\tau_0,\nu)\not=0$ for a given $\nu$.

\begin{figure}[t]
\begin{center}
\hspace*{2mm}\includegraphics[width=0.95\linewidth]{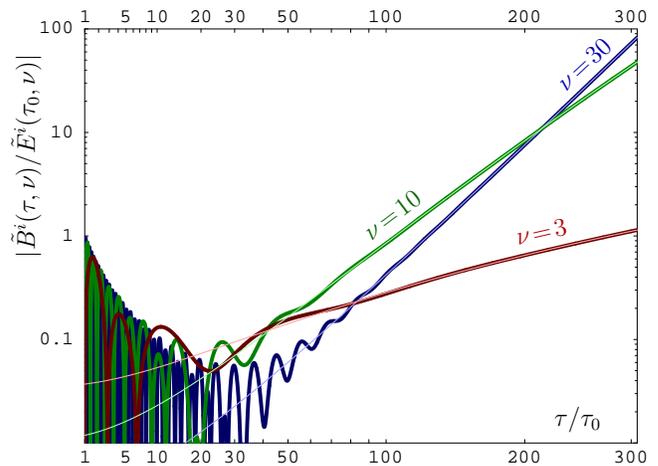}
\vspace*{-4mm}
\end{center}
\setlength{\unitlength}{1cm}
\caption{Numerical solution for individual magnetic modes
$\widetilde{B}^i(\tau,\nu)$ normalized to 
$\widetilde{E}^i(\tau_0,\nu)=\tau_0^{-1} \widetilde{\Pi}^i(\tau_0,\nu)$
(thick dark lines) for $\tau_0/\tau_{\rm iso}=100$ and $c=0.5$
in a log-square root plot.
Thin light lines give
the analytical result of
Eq.(\ref{intermediate})  with normalization
to match the numerical result at late times.
Note that $\tau_0^{-1}\simeq Q_s \simeq 1$ and 3 GeV for
RHIC and LHC, resp.\ \cite{Iancu:2003xm}.}
\label{fig:comparison}
\end{figure}

In order to fix our dimensionful parameters in a way
that makes contact with heavy-ion physics, we adopt the
saturation scenario \cite{McLerran:1993ni,Iancu:2003xm} and assume that at $\tau_0\simeq
Q_s^{-1}$ we have an initial hard-gluon density \cite{Baier:2002bt}
$n(\tau_0)=2 c Q_s^3/(3\pi^2 \alpha_s Q_s \tau_0)$, where $Q_s$ is the
saturation scale. 
We consider two cases for the gluon liberation factor $c$:
$c\simeq 0.5$ according to numerical simulations of Ref.~\cite{Krasnitz:1998ns}
and $c=2\ln2$ according to an approximate analytical calculation
of Ref.~\cite{Kovchegov:2000hz}. Assuming further a deformed thermal
distribution $f_0$ with $\tau_{\rm iso}\le \tau_0$ and
transverse temperature $T=Q_s/d$ with $d^{-1}=0.47$ taken from Ref.~\cite{Iancu:2003xm},
we obtain for a purely gluonic system
$\mu/Q_s=cd\pi^2/[48\zeta(3)]\approx 0.182$
for $c=0.5$ and 
$\approx 0.505$ for
$c=2\ln 2$. The amount of anisotropy at time $\tau_0$ is
determined by the ratio $\tau_0/\tau_{\rm iso}$.

In Fig.~\ref{fig:comparison} we display our numerical results
for the magnetic field strength 
$\widetilde{B}^i(\tau,\nu)=\nu \tau^{-1}\, \widetilde{A}^i(\tau,\nu)$
with $c=0.5$ 
for $\nu=3,10,30$ and high initial
anisotropy, $\tau_0/\tau_{\rm iso} = 100$ (i.e., $\xi(\tau_0)\sim 10^4$). 
For late times we observe a nearly perfect agreement with
the analytical estimate of Eq.(\ref{intermediate}), which however
does not contain information on the amplitudes of the
unstable modes compared to their initial values.
In the numerical evaluation we find that the modes
with larger longitudinal momentum $\nu$, 
which have larger
growth rates at late times, typically start to grow also only at larger
proper times. This causes a certain delay for the onset
of the instability which can be measured, e.g., by the time
$\tau_{10}$ when the first of the modes 
$\widetilde{A}_i$
has grown in amplitude by a factor of 10
relative to its first maximum.
For $\tau\ge\tau_{10}$ the dynamics of the collective modes
is clearly dominated by the unstable modes. For times $\tau$ up to 
about half of $\tau_{10}$ we observe a decrease
of the energy density carried by the unstable modes due to the expansion,
but increase despite continued expansion thereafter.
In Table~\ref{tablem}
we list the values of $\tau_{10}$ that we found by scanning
through $\nu$ with various
initial anisotropies. 
Ideal-hydrodynamic fits to experimental data at RHIC indicate a life-time 
of a quark-gluon plasma of less than 5 fm/c $\simeq30\, \tau_0$ 
\cite{Huovinen:2006jp}.
The values obtained in Table~\ref{tablem} turn out
to be too large to suggest an important role of plasma instabilities
in RHIC experiments if the gluon liberation factor $c\simeq0.5$ 
as obtained in Ref.~
\cite{Krasnitz:1998ns,Baier:2002bt}. 
However, these unstable modes could
contribute to fast isotropization of a quark-gluon plasma if the 
value of $c$ is much larger than those currently considered, or if the 
viscosity of the plasma is significant (in which case the life-time is 
increased \cite{Baier:2006um}). Finally, our results suggest that even for 
$c\simeq 0.5$, plasma instabilities 
will be an important 
phenomenon at the LHC, where plasma life-times might exceed 
7 fm/c$\,\simeq100\, \tau_0$ \cite{Eskola:2005ue}.

\begin{table}
\caption{Approximate values of the proper time $\tau_{10}$ where
the first 
of the modes $\widetilde{A}^i(\tau,\nu)$ 
has grown by a factor 10.
\label{tablem}}
\begin{ruledtabular}
\begin{tabular}{l|rrrr}
$\tau_0/\tau_{\rm iso}=\sqrt{\xi(\tau_0)+1}$ & 1 & 10 & 100 & 1000 \\
\hline
$\tau_{10}/\tau_0$ ($c=0.5$)    &260&60&50&49\\
$\tau_{10}/\tau_0$ ($c=2\ln 2$) &95  &25&21&20  \\
\end{tabular}
\end{ruledtabular}
\end{table}

To conclude, in this Letter we have laid the basis for
numerical simulations of 
non-Abelian dynamics
in heavy-ion collisions
by generalizing the anisotropic but stationary
hard-loop effective theory to the 
longitudinally expanding case. 
This opens the possibility of 
studying the fate of non-Abelian
plasma instabilities in the highly nonlinear regime
with expansion included. 
Moreover, we have determined the evolution of Weibel instabilities
in a longitudinally expanding plasma in the regime where
non-Abelian self-interactions of the unstable modes are negligible,
finding that they start out oscillatory
and later grow fast with an asymptotic behavior that
we could determine analytically.
Numerically, we
have also been able to quantify the onset of this instability. 
Matching our dimensionful parameters to those of the saturation
scenario we find that plasma instabilities overcome the effects of expansion
at or after
$\tau\sim 20 Q_s^{-1}$ (which is roughly consistent with 
Ref.\cite{Romatschke:2005pm}). 
Unless initial parton densities are significantly higher than assumed here,
only the prospected LHC experiments seem to offer 
large enough $Q_s$ and plasma life-times 
to generate strong
quark-gluon-plasma instabilities from
small rapidity fluctuations.
The dynamics of strong initial fluctuations can only be determined
by full nonlinear studies; this is work in progress.

\acknowledgments

We are grateful to Mike Strickland for collaboration and 
stimulating discussions.
PR was supported by BMBF 06BI102.


\end{document}